\documentclass[pre,aps,preprint,showpacs]{revtex4}
\usepackage{revsymb,latexsym,amssymb,amsmath,comment}
\usepackage{graphicx,psfrag,subfigure,stmaryrd}

\begin{document}
\author{A. Golovnev, S. Trimper} 
\affiliation{Institute of Physics,
Martin-Luther-University, D-06099 Halle Germany}
\email{anatoly.golovnev@physik.uni-halle.de;steffen.trimper@physik.uni-halle.de}
\title{Non-linear effects in electrolytes at large applied voltage}
\date{\today }

\begin{abstract}
\noindent The steady state of ions diffusion in polymer electrolytes at arbitrary applied voltage is analyzed in 
the framework of the Nernst-Planck-Poisson equation (NPP). The exact solution of the set of equations is found 
without the assumption of low ions concentration. The solution is independent of the kinetic properties of the 
system. At constant voltage there is a master curve for concentration in terms of the initial concentration. 
Enhancing the voltage causes  an increase of the ion concentration gradient and consequently the applicability 
of the NPP is violated for high voltages. The analytical finding is estimated by using experimental data from 
recent measurements (P. Kohn et al Phys. Rev. Left. {\bf 99}, 086104 (2007)). As the result we find an upper 
bound for the validity of the NPP. Above this voltage higher order gradient terms become relevant.

\end{abstract}

\maketitle

\section{Introduction}

\noindent Polymer electrolytes characterized as amorphous structure are widely used for solid state 
ionics devices such as lithium ion batteries \cite{ta}, fuel cells, electrochemical displays etc. The 
material is a composite one consisting of dissociated ions which are dissolved in a polymer or glassy 
matrix. Very recently the motions of proteins in densely grafted layers of polyelectrolyte brushed has been 
studied \cite{he} or the dynamics of polyelectrolyte transport through a protein channel as a function 
of applied voltage \cite{br}. The understanding of the ionic transport mechanism in polymer materials is 
of basic interest and has attracted attentions over several decades, for a recent review on the history, 
the applications of ionic transport and structure as well as dynamics see \cite{ka}. A number of different 
theoretical methods has been proposed for polyelectrolyte systems \cite{a,d,p}. The activities concerning 
the theoretical approach for the ionic transport behavior had been reviewed in \cite{1}. 

\noindent Our analysis is based on the Nernst-Planck equations supplemented by the static Poisson equation. 
These set of equations, abbreviated as NPP, are derived for a low charge carrier density. The diffuse charge 
dynamics is characterized by a linear gradient term for the ionic concentration. In addition to the diffusive 
charge dynamics the transport process is driven by an external field $\vec E$. For low voltages the linearized 
NPP can be solved by Laplace transforms \cite{1} whereas for high voltages numerical solutions had been 
obtained in \cite{1}. As estimated in \cite{1} the validity of the linear approximation should be guaranteed 
up to 70 mV. A detailed analysis of the non-linear regime is still lacking. Otherwise, the mobility of ions 
has been determined recently by transient current measurements at high voltages up to the order of 300 V \cite{4}. 
The problem arisen is whether the NPP with a linear concentration gradient are also valid for such high fields. 
Intuitively one expects that high voltages should give rise to higher order concentration gradients of the mobile 
ions. Otherwise, very low voltages cause only low gradients even in systems with a high concentration of ions.
Thus, the limitation of the NPP is related to both high concentration of charge carriers as well high applied 
voltages. However our analysis offers that a high initial concentration should be irrelevant. Whereas all previous 
papers are mainly devoted to a numerical solution of the NPP or some modifications of those equations such as by 
including steric effects \cite{2}, the present paper is directly addressed to the NPP, in order to check the ability 
of the equations at high voltages. Despite the model is used widely it seems to be difficult to estimate 
the condition under which the NPP are valid. So far, the NPP has not been solved exactly but instead of that 
the solution had been discussed in a linear regime where the long time behavior was found. Here we present 
the exact steady state solution of the full NPP. Especially we demonstrate that the solution within the linear 
regime may be spread up to $70$ mV approximately. Otherwise the experimental measurement of the mobility is fulfilled 
in between 70 and 300 V \cite{4}. Obviously the NPP is not appropriate to clarify the observations made in \cite{4}. 
Apparently, a high voltages requires higher order diffusive terms. 

\section{The model and the solution}

\noindent Let us consider a solution of charged particles embedded in polymer electrolyte at temperature above the 
glass-transition temperature of the polymer. We study completely dissociated electrolyte 
placed in between infinite flat electrodes at coordinates $x= -L$ and $x = L$. Therefore the system offers a simple 
one-dimensional geometry. Concentrations of the charged ions are described by continuum fields $C_{\pm}(x,t)$. 
The charged ions are subjected to an electric field $\vec E = - \nabla \phi$. The scalar potential $\phi $ 
obeys the Poisson equation
\begin{equation}\label{key1}
-\epsilon\frac{\partial^{2}\phi}{\partial x^{2}}= ze\left( C_{+}-C_{-}\right) \,,
\end{equation}
\noindent where $\epsilon$ is the dielectric permeability, $e$ is the elementary charge and $ze$ is the charge of the ion. 
Because the charge is conserved the time derivative of $C_{\pm}$ is given by the spatial 
derivative of the corresponding ionic fluxes denoted by $F_{\pm}$. Generally this flux is composed of two terms, 
namely the diffusive part originated by a small concentration gradient and the electric field term. This field 
is assumed to couple linearly to the concentration field. The set of evolution equations reads
\begin{eqnarray}\label{key2}
\frac{\partial C_{+}}{\partial t}&=& -\frac{\partial}{\partial x}\left(F_{+}\right)=-\frac{\partial}{\partial x}
\left(-D\frac{\partial C_{+}}{\partial x}-\mu zeC_{+}\frac{\partial\phi}{\partial x} \right)\nonumber\\ 
\frac{\partial C_{-}}{\partial t}&=&-\frac{\partial}{\partial x}\left(F_{-}\right)=
-\frac{\partial}{\partial x}\left(-D\frac{\partial C_{-}}{\partial x}+\mu zeC_{-}\frac{\partial\phi}{\partial x} \right)\,.
\end{eqnarray}
\noindent Here $D$ is the diffusivity and $\mu$ denotes the mobility, for simplicity we have assumed 
$D_{+}=D_{-}=D$. Further it is supposed that the Einstein relation $D=\mu kT$ is fulfilled. Likewise it is 
supposed that the Faraday currents can be neglected. The boundary conditions imposed are the 
disappearance of fluxes $F_{\pm}=0$ at $x=\pm L$. This set of equations combined with Eq.~\eqref{key1} are 
called Nernst-Planck-Poisson equations (NPP). Essentially is that Fick's law of diffusion is fulfilled, i.e.  
the ionic flux is proportional to the first derivative of concentration. At high voltage, the gradient of 
concentration becomes more pronounced and the consecutive equations should be changed due to higher order gradient 
terms. To estimate the validity of the NPP for high voltages let us consider the steady state. This state is 
characterized by constant ionic fluxes where the constant is zero due to the boundary conditions. For that case 
we found an exact solution demonstrated further. The difference of the charge carriers due Eqs.~\eqref{key2} 
can be expressed by the potential of the electric field according to Eq.~\eqref{key1}. It results
\begin{equation}
\frac{D\epsilon}{ze} \frac{\partial^3 \phi(x)}{\partial x^3} - \mu ze \frac{\partial \phi(x)}{\partial x} 
C(x)=0\quad {\rm with}\quad C(x) = C_+(x) + C_-(x)\,.
\label{key2a}
\end{equation}
Using Eqs.~\eqref{key2} and \eqref{key1} the sum of the concentration $C(x)$ satisfies  
\begin{equation}
-D C(x) + \frac{\mu \epsilon }{2} \left(\frac{\partial \phi(x)}{\partial x}\right)^2 = r\,,
\label{key2b}
\end{equation}
where $r$ is an integration constant.
Both Eqs.~\eqref{key2a} and \eqref{key2b} can be combined to a single equation for the electric potential
\begin{equation}
\frac{D\epsilon}{ze}\frac{\partial^3 \phi}{\partial x^3} - \frac{\mu ze}{D}\frac{\partial \phi}{\partial x}
\left[\frac{\mu\epsilon}{2}\left( \frac{\partial \phi}{\partial x}\right)^{2}-r\right] = 0\,.
\label{key3}
\end{equation}
Then the solution of the last equation can be applied to find the charge concentration via
\begin{equation}
\frac{\partial \ln C_{\pm }(x)}{\partial x} = \mp \frac{\mu z e}{D} \frac{\partial \phi(x)}{\partial x}
\label{key3a}
\end{equation}
Actually the solution of Eq.~\eqref{key3} is given in terms of Jacobian elliptic functions \cite{3}. 
To that aim let us substitute $y(x) = \phi'(x)$ in Eq.~\eqref{key3}. After some formal steps we end up 
with an expression for the electric field $E(x)$
\begin{equation}
E(x) = - \frac{d \phi(x)}{dx} = -\frac{2D}{\mu ze}\cdot q\cdot\beta\cdot sn(\beta x-x_{0},q)\,.
\label{key5}
\end{equation}
Using Eq.~\eqref{key3a} and the properties of the Jacobian functions \cite{3} we get the stationary 
concentration profiles of the charge carriers 
\begin{equation}\label{key6}
C_{\pm}(x) =-\frac{D\epsilon}{\mu z^{2}e^{2}}\cdot q\cdot\beta^{2}\cdot
\left[dn(\beta x-x_{0},q)\pm q\cdot cn(\beta x-x_{0},q) \right] ^{2}\,.
\end{equation}
\noindent The quantities $sn$, $cn$ and $dn$ are the Jacobian elliptical functions, which are 
characterized by two parameters denoted as $\beta$ and $q$. Due to the symmetry property 
$E(x)=E(-x)$ the integration constant $x_{0}$ should be equal to quarter of the period of Jacobian sinus 
$x_{0}=K(q)$, where $K(q)$ is the complete elliptic integral of the first kind \cite{3}.  
The parameters $\beta$ and $q$ are determined by the following conditions:
\begin{eqnarray}\label{key8}
\int^L_{-L} C_{\pm} (x)\cdot dx &=& 2L \eta \nonumber\\
E({\pm} L)=-\frac{V}{2L}\,.
\end{eqnarray}
The first condition determines the total concentration $\eta$ which is given initially. Thus, $\eta$ 
is the initial particle concentration before the voltage were applied. The second condition fixes the 
applied voltage $V$. After performing the integrating of the first equation in Eq.~\eqref{key8} 
and reorganization of the the second one we have
\begin{eqnarray}\label{key10}
E(\beta L,q)-q^{2}sn(\beta L,q)cd(\beta L,q)&-&\frac{\beta L}{2}\left(1-q^{2}\right)=
-\frac{L\eta}{2\beta q}\frac{\mu z^{2}e^{2}}{D\epsilon}\nonumber\\ 
cd(\beta L,q)&=& -\frac{V}{L\beta q}\frac{\mu ze}{4D}\,.
\end{eqnarray}
Here $E(\beta L,q)$ stands for the incomplete elliptic integral of the second kind and we have introduced 
$$
cd(\beta x,q)\equiv \frac{cn(\beta x,q)}{dn(\beta x,q)}\,.
$$
Eqs.~\eqref{key5}, \eqref{key6} together with Eqs.~\eqref{key10} provide the complete solution of the stationary 
problem of the NPP. However the conditions in Eqs.\eqref{key10} for the parameters $\beta$ and $q$ are too 
complicated to get analytical results. To proceed one can use the fact that in the first equation the first term  
is slowly changing almost linear function, whereas the second term can be roughly approximated by just $q$. 
This approximation leads to an parameter area which enables us to get reasonable values for the parameters. 

\section{Results}

\noindent All calculations were performed for a real system described in \cite{4}. The experiments had been 
carried in thin films with $L = 5\cdot 10^{-5}$ m. The initial concentration $\eta$ 
is varied from $1\cdot 10^{21}$ to $1.3\cdot 10^{24}$ $m^{-3}$, whereas the applied voltage $V$ was changed from 
$1\cdot 10^{-3}$ to $2.6\cdot 10^{-1}$ V. In case the parameter $q$ of the Jacobian elliptic function is 
restricted to the internal $0 \leq q \leq 1$, the functions are real. Inserting the experimental data obtained in 
\cite{4} in our solution we find $q >> 1$ and $\beta << 1$. In particular it results $q \simeq 10^{9}$ and 
$\beta \simeq 10^{-5}$. For such values of the parameters $q$ and $\beta$ the Jacobian functions become complex 
functions. However it appeared that $\Im E << \Re E$ and $\Im C_{\pm } << \Re C_\pm $. We estimate that 
the imiginary part is about 7 or 8 orders of magnitude smaller than the real part. Despite of the complexness of 
the Jacobian functions, the solution found in the paer has a physical sense.

\subsection{Results for constant voltage}

\begin{figure}[ht]
\includegraphics[angle=-90,width=1.0\textwidth]{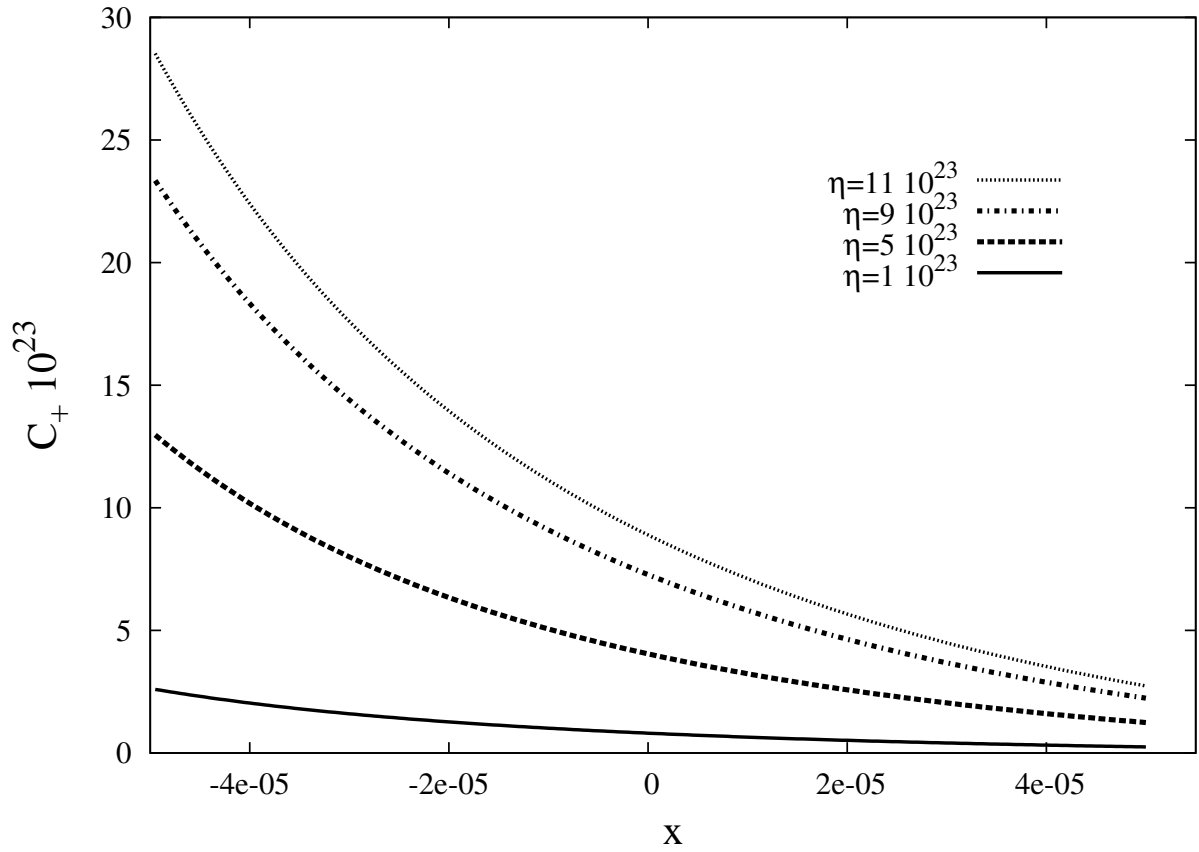}
\caption{Concentration profile $C_{+}(x)$ in the capacitor. Different curves 
correspond to different initial concentrations $\eta$. 
The applied voltage $V$ is assumed to be $5\cdot 10^{-2}$V.}
\label{Fig.1a}
\end{figure}

\begin{figure}[ht]
\includegraphics[angle=-90,width=1.0\textwidth]{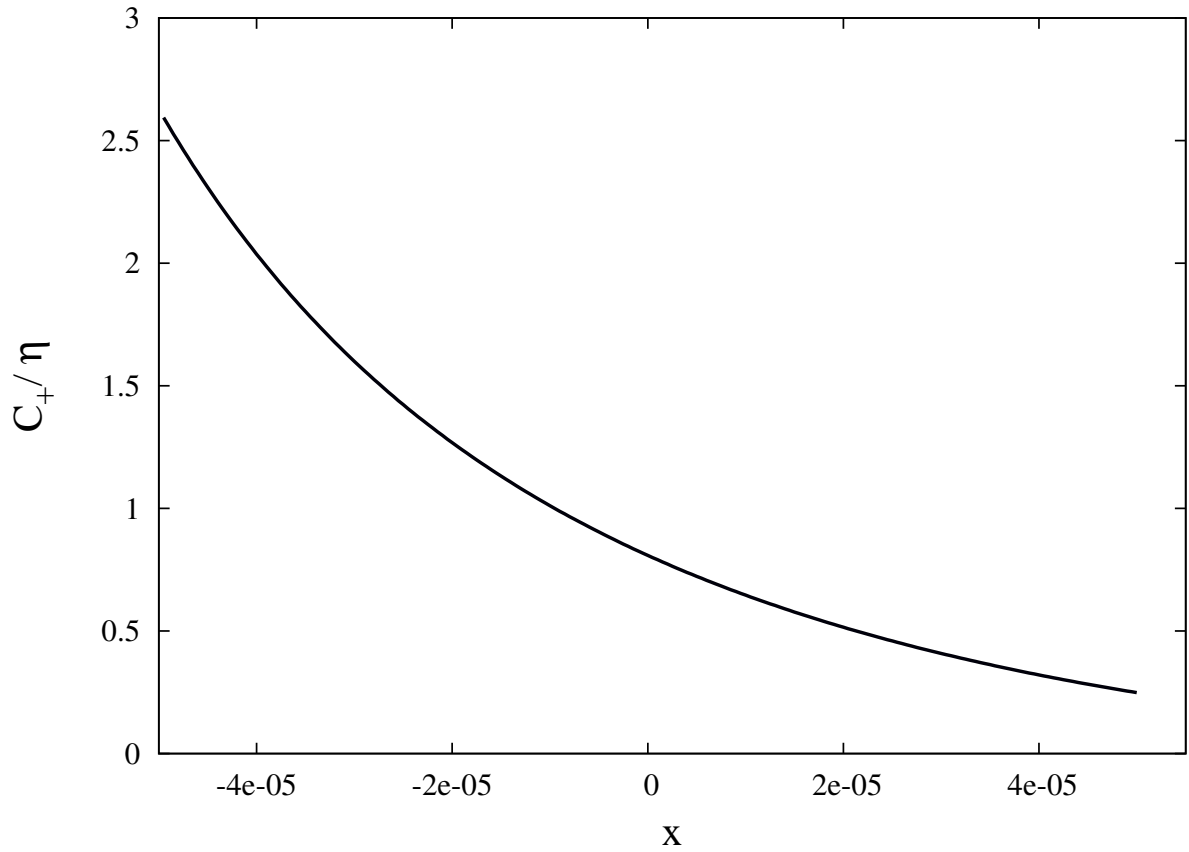}
\caption{Concentration profile $\frac{C_{+}(x)}{\eta}$ in terms of the initial concentration $\eta$ in the capacitor. 
The applied applied voltage is $ V = 5\cdot 10^{-2}$V.}
\label{Fig.1b}
\end{figure}

\begin{figure}[ht]
\includegraphics[angle=-90,width=1.0\textwidth]{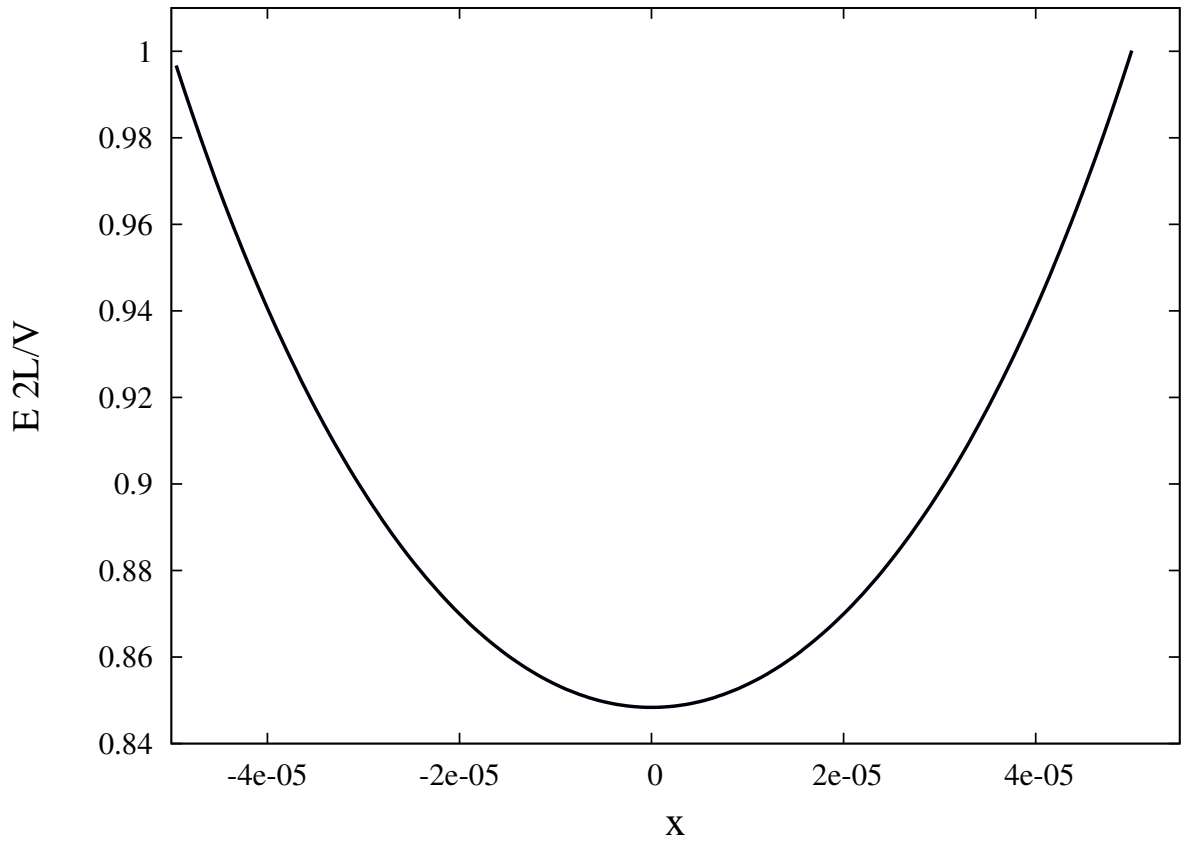}
\caption{Electric field profile $E(x)$ within the capacitor. The profile is independent  on the initial 
concentration $\eta$. The voltage is assumed to be $5\cdot 10^{-2}$V.}
\label{Fig.2}
\end{figure}

\noindent Calculations at constant voltage $V$, defined in Eq.~\eqref{key8}, offer that the parameter 
$q$ changes proportional to the initial concentration $\eta$, whereas the other parameter$\beta$ behaves 
proportional to $\eta^{-1}$, i. e. we can set $\beta=\frac{\tilde \beta}{\eta}$ and $q=\tilde q \eta$, 
where the tilde means a value at some definite concentration. Notice that for different applied voltage $V$ 
the parameters $\tilde q$ and $\tilde \beta$ are also different. As one can see from Eq.~\eqref{key8}, 
the integral $\int^L_{-L} \frac{C_{\pm} (x)}{\eta}\cdot dx $ is independent on the fixed initial concentration 
$\eta$ indicating the existence of a master curve for $\frac{C (x)}{\eta}$. Our calculation confirms the 
presumption. In Fig.\ref{Fig.1a} the spatial charge distribution 
is shown according to Eq.~\eqref{key6}. The different curves depicted in Fig.~\ref{Fig.1a} correspond to different 
values for $\eta$. In Fig.\ref{Fig.1b} one observes that all these data for the concentration in terms of $\eta$ lay 
on a single master curve. Further discussions concerning scaling properties of the curves one can in \cite{5} . 
The profile of the electric field is independent of the concentration which is demonstrated in Fig.~\ref{Fig.2}, with other 
words the efficiency of screening is independent of $\eta$. This observation would suggest the applicability of NPP 
is guaranteed also for a high charge carrier concentration. According to Eq.~\eqref{key3a} a constant electric field 
suggest an increasing of the concentration gradient proportional to the concentration itself. Therefore high gradients 
could lead to an inconsistency of NPP.
\noindent Another feature mentioned is the so called charge inversion, i.e. interfacial charges attract counterions 
in excess of their own nominal charge. This effect were found both in theoretical \cite{6} and experimental \cite{7,8} 
works; for an overview and discussion see \cite{9}. But, as one can see from Fig.\ref{Fig.1b}, in the framework of 
steady state solution of NPP this effect is absent.

\subsection{Results for constant concentration}

\begin{figure}[ht]
\includegraphics[angle=-90,width=1.0\textwidth]{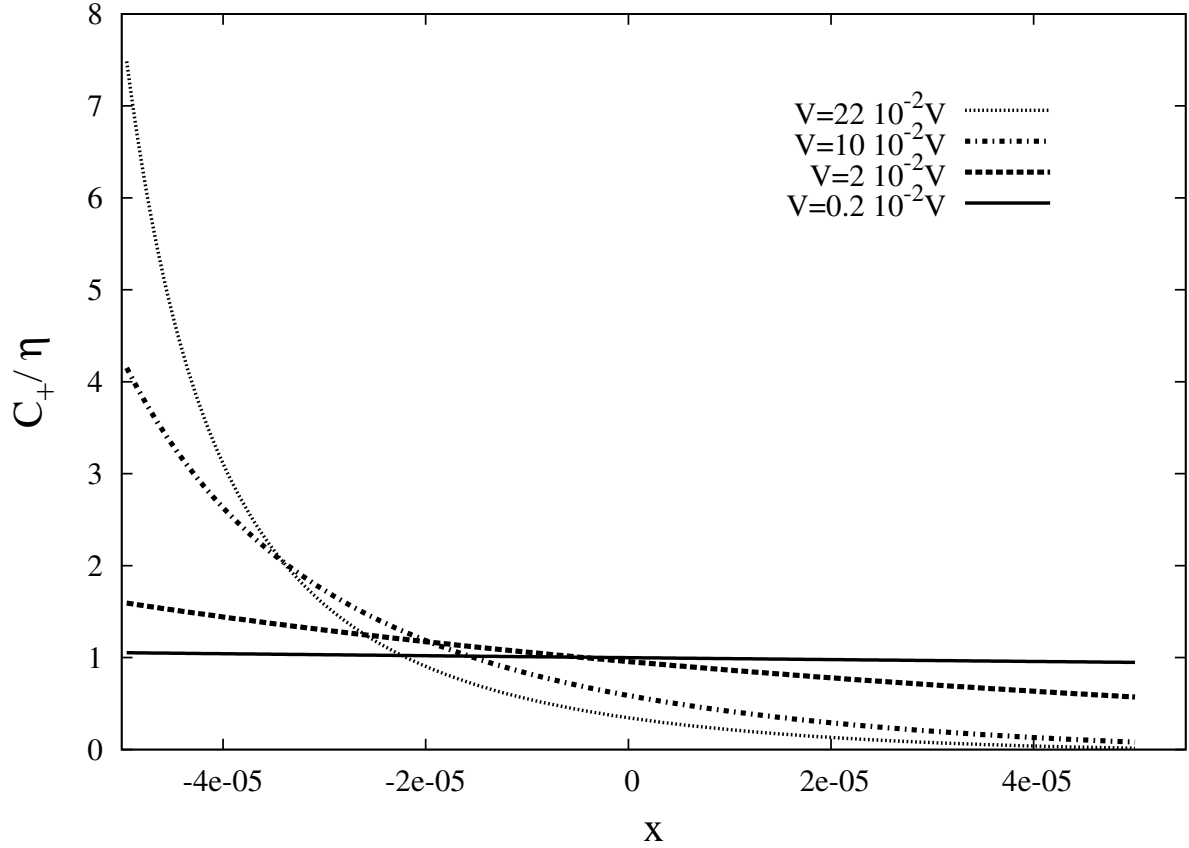}
\caption{Concentration profile $C_{+}(x)$ versus coordinate in the capacitor. Different curves correspond 
to different applied voltages $V$. The initial concentration $\eta$ is $5\cdot 10^{23}$.}
\label{Fig.3}
\end{figure}

\noindent For different voltages but constant concentration there exists no master curve as it is shown in 
Fig.~\ref{Fig.3}. For increasing applied voltage $V$ we find oscillations of the electric field $E(x)$. From 
here we conclude that the NPP equations are not longer adequate in describing the physical situation in mind. 
Such oscillations appear when the period of the Jacobian sinus is sufficiently small. Analytically the criterion 
for the absence of oscillations is the condition $\beta L\le K(q)$. In case of $\beta L = K(q)$ the half 
period of the Jacobian sinus fits exactly the capacitor between $-L \leq x \leq L$. To be more specific we have 
checked that for an arbirtrary but fixed initial concentration $\eta$ the inequality $\beta L < K(q)$ is only 
satisfied in case the applied voltages fulfills $V \leq 0.22$ V. Because for $V \geq 0.23$ V the inequality is 
violated we conclude that there exist a kind of critical applied voltage in between $0.22 \leq V \leq 0.23$ V. 
If any the dependence of that result on the initial concentration is weak. But such a weak dependence 
should exist due to Eq.~\eqref{key3a}, which signalized that the concentration gradient is proportional to both 
the applied voltage $V$ and the initial concentration $\eta$. 

\noindent The steady state solution of NPP equations is universal and nearly independent on specific system 
properties. Thus the kinetic properties should depend on the diffusion coefficient $D$. But in the expression 
for the steady state according to Eq.~\eqref{key6} there appears only the ratio $D/\mu$. If Einstein relation 
is fulfilled the solution is independent on the kinetic properties of the system but depends only on the temperature. 
In other words, different systems with the same $ze$ tend to the same steady state solution, which is however 
reached on different ways. If the voltage $V$ is sufficiently high to alter the transport properties of polymers 
(mechanisms for a change are discussed in \cite{10}) but not high enough to alter its dielectric permeability, 
the steady state will be unchanged. 

\section{Conclusion}

\noindent By getting the steady state we found a limit of the applicability of the Poisson-Nernst-Planck equations 
for both high voltages and high concentrations. Using the experimental data from \cite{4} we could estimate that 
above $V = 0.22$ V the NPP yield unphysical results characterized by the appearance of unphysical oscillation of the 
solutions. Notice that the validity of the linear approximation is restricted to an applied voltage 
$V \leq 0.07$ V \cite{1}. The limit of the applicability of NPP due to a high concentration of ions is not 
very sharp. Unlike a behavior at high voltage, in this case NPP do not contain explicit restrictions itself. 
Existents of the master curve for normalized concentration is an indication for that fact. The inclusion of higher 
order gradient terms leads a more complicated model. Otherwise the assumption that the flux is proportional to a 
linear gradient term is a very promising assumption which should be valid for the majority of situations.\\

\noindent One of us (A. G.) acknowledges support by the International Max Planck Research 
School for Science and Technology of Nanostructures in Halle, Germany and Thomas Michael 
for helping in preparing the figures. Further we thank Peter Kohn 
for useful discussions.

\newpage

\end{document}